\documentclass[a4paper]{amsart}

\RequirePackage{amsmath}
\RequirePackage{bm}
\RequirePackage{amssymb}
\RequirePackage{upref}
\RequirePackage{amsthm}
\RequirePackage{enumerate}
\RequirePackage{pb-diagram}
\RequirePackage{amsfonts}
\RequirePackage[mathscr]{eucal}
\RequirePackage{verbatim}
\RequirePackage{xr}
\RequirePackage{graphicx}
\usepackage{calc}
\usepackage{xspace}
\RequirePackage{color}
\RequirePackage{ifthen}
\RequirePackage{fancybox}
\RequirePackage{mathrsfs}

\newcommand{\cf}{cf.\@\xspace}
\newcommand{\resp}{resp.\@\xspace}

\newcommand{\al}{\alpha}

\newcommand{\ga}{\gamma}
\newcommand{\de}{\delta }
\newcommand{\e}{\epsilon}

\newcommand{\f}{\varphi}

\newcommand{\lam}{\lambda}
\newcommand{\m}{\mu}

\newcommand{\s}{\sigma}

\newcommand{\D}{\varDelta}
\newcommand{\F}{\varPhi}
\newcommand{\Lam}{\varLambda}
\newcommand{\Om}{\varOmega}

\newcommand{\so}{{\mc S_0}}

\newcommand{\msp[1]}[1]{\mspace{#1mu}}

\newcommand{\R}[1][n+1]{{\protect\mathbb R}^{#1}}

\newcommand{\Cc}{{\protect\mathbb C}}

\newcommand{\N}{{\protect\mathbb N}}

\newcommand{\eR}{\stackrel{\lower1ex \hbox{\rule{6.5pt}{0.5pt}}}{\msp[3]\R[]}}
\newcommand{\eN}{\stackrel{\lower1ex \hbox{\rule{6.5pt}{0.5pt}}}{\msp[1]\N}}
\newcommand{\eO}{\stackrel{\lower1ex \hbox{\rule{6pt}{0.5pt}}}{\msc O}}

\newcommand\ra{\rightarrow}

\newcommand\hra{\hookrightarrow}

\newcommand{\un}{\infty}
\newcommand{\A}{\forall}

\newcommand{\set}[2]{\{\,#1\colon #2\,\}}
\newcommand{\uu}{\cup}
\newcommand{\ii}{\cap}
\newcommand{\uuu}{\bigcup}
\newcommand{\iii}{\bigcap}
\newcommand{\uud}{ \stackrel{\lower 1ex \hbox {.}}{\uu}}
\newcommand{\uuud}[1]{ \stackrel{\lower 1ex \hbox {.}}{\uuu_{#1}}}
\newcommand\su{\subset}

\newcommand{\sminus}[1][28]{\raise 0.#1ex\hbox{$\scriptstyle\setminus$}}

\newcommand{\wh}{\widehat}

\newcommand{\wed}{\wedge}

\newcommand{\abs}[1]{\lvert#1\rvert}

\newcommand{\norm}[1]{\lVert#1\rVert}

\newcommand{\spd}[2]{\protect\langle #1,#2\protect\rangle}

\newcommand{\tit}{\textit}

\newcommand{\tup}{\textup}

\newcommand{\mc}{\protect\mathcal}
\newcommand{\msc}{\protect\mathscr}

\providecommand{\bysame}{\makebox[3em]{\hrulefill}\thinspace}

\newcommand{\bt}{\begin{thm}}
\newcommand{\bl}{\begin{lem}}
\newcommand{\bc}{\begin{cor}}
\newcommand{\bd}{\begin{definition}}
\newcommand{\bpp}{\begin{prop}}
\newcommand{\br}{\begin{rem}}
\newcommand{\bn}{\begin{note}}
\newcommand{\be}{\begin{ex}}
\newcommand{\bes}{\begin{exs}}
\newcommand{\bb}{\begin{example}}
\newcommand{\bbs}{\begin{examples}}
\newcommand{\ba}{\begin{axiom}}
\newcommand{\bas}{\begin{assumption}}

\newcommand{\et}{\end{thm}}
\newcommand{\el}{\end{lem}}
\newcommand{\ec}{\end{cor}}
\newcommand{\ed}{\end{definition}}
\newcommand{\epp}{\end{prop}}
\newcommand{\er}{\end{rem}}
\newcommand{\en}{\end{note}}
\newcommand{\ee}{\end{ex}}
\newcommand{\ees}{\end{exs}}
\newcommand{\eb}{\end{example}}
\newcommand{\ebs}{\end{examples}}
\newcommand{\ea}{\end{axiom}}
\newcommand{\eas}{\end{assumption}}

\newcommand{\bp}{\begin{proof}}
\newcommand{\ep}{\end{proof}}
\newcommand{\eps}{\renewcommand{\qed}{}\end{proof}}

\newcommand{\bal}{\begin{align}}

\newcommand{\bi}[1][1.]{\begin{enumerate}[\upshape #1]}
\newcommand{\bia}[1][(1)]{\begin{enumerate}[\upshape #1]}
\newcommand{\bin}[1][1]{\begin{enumerate}[\upshape\bfseries #1]}
\newcommand{\bir}[1][(i)]{\begin{enumerate}[\upshape #1]}
\newcommand{\bic}[1][(i)]{\begin{enumerate}[\upshape\hspace{2\cma}#1]}
\newcommand{\bis}[2][1.]{\begin{enumerate}[\upshape\hspace{#2\parindent}#1]}
\newcommand{\ei}{\end{enumerate}}

\newcommand\ndots{\raise 0.47ex \hbox {,}\hskip0.06em\cdots %
     \raise 0.47ex \hbox {,}\hskip0.06em} 

\newcommand{\q}{\quad}
\newcommand{\qq}{\qquad}

\newcommand\nd{\noindent}

\newskip\Csmallskipamount                                                
\Csmallskipamount=\smallskipamount
\newskip\Cmedskipamount
\Cmedskipamount=\medskipamount
\newskip\Cbigskipamount
\Cbigskipamount=\bigskipamount

\newcommand\cvs{\vspace\Csmallskipamount}   
\newcommand\cvm{\vspace\Cmedskipamount}

\newskip\csa
\csa=\smallskipamount

\newskip\cma
\cma=\medskipamount

\newskip\cba
\cba=\bigskipamount

\newdimen\spt
\newcommand\citem{\cvs\advance\itemno by
1{(\romannumeral\the\itemno})\hskip3pt}
\newcommand{\bitem}{\cvm\nd\advance\itemno by
1{\bf\the\itemno}\hspace{\cma}}

\newcount\itemno
\newcommand{\lae}[1]{\label{E:#1}}
\newcommand{\lat}[1]{\label{T:#1}}
\newcommand{\lal}[1]{\label{L:#1}}

\newcommand{\laas}[1]{\label{Ass:#1}}

\newcommand{\rt}[1]{Theorem~\ref{T:#1}}
\newcommand{\rl}[1]{Lemma~\ref{L:#1}}

\newcommand{\ras}[1]{Assumption~\ref{Ass:#1}}

\newcommand{\re}[1]{\eqref{E:#1}}

\newcommand{\frt}[1]{Theorem~\ref{T:#1} on page~\tup{\pageref{T:#1}}}

\newcommand{\fre}[1]{\eqref{E:#1} on page~\tup{\pageref{E:#1}}}

\newcommand{\fras}[1]{Assumption~\ref{Ass:#1} on page~\tup{\pageref{Ass:#1}}}

\newskip\thmskip
\thmskip=\parindent

\newskip\hsk
\newenvironment{hinw}{\labelsep=0pt\begin{list}{}{\labelsep=0pt\itemindent=0pt\labelwidth=0pt\leftmargin=\parindent\rightmargin=0pt\partopsep=\cba}%
\item\it\nopagebreak\nopagebreak}%
{\end{list}}

\newcommand\bh{\begin{hinw}}
\newcommand{\eh}{\end{hinw}}

\newtheoremstyle{normal}
  {\cba}
  {\cba}
  {}
  {\thmskip}
  {\bfseries}
  {.}
  {\hsk}
  {}

\newtheoremstyle{abschnitt}
  {\cba}
  {\cba}
  {}
  {\thmskip}
  {\bfseries}
  {.}
  {\hsk}
  {}

\newtheoremstyle{italic}
  {\cba}
  {\cba}
  {\itshape}
  {\thmskip}
  {\bfseries}
  {.}
  {\hsk}
  {}

\newtheoremstyle{aufgaben}
  {\cba}
  {\cba}
  {}
  {}
  {\normalsize\bfseries}
  {.}
  {\hsk}
  {}

\newtheoremstyle{break}
  {\cba}
  {\cba}
  {\itshape}
  {}
  {\bfseries}
  {.}
  {\newline}
  {}

\swapnumbers
\theoremstyle{italic}
\newtheorem{thm}[subsection]{Theorem}
\newtheorem{lem}[subsection]{Lemma}
\newtheorem{prop}[subsection]{Proposition}
\newtheorem{cor}[subsection]{Corollary}

\theoremstyle{normal}
\newtheorem{rem}[subsection]{Remark}
\newtheorem{definition}[subsection]{Definition}
\newtheorem{example}[subsection]{Example}
\newtheorem{examples}[subsection]{Examples}
\newtheorem{ex}[subsection]{Exercise}
\newtheorem{note}[subsection]{}
\newtheorem{axiom}[subsection]{Axiom}
\newtheorem{assumption}[subsection]{Assumption}

\theoremstyle{aufgaben}
\newtheorem{exs}[subsection]{Exercises}

\swapnumbers

\numberwithin{equation}{section}
\numberwithin{figure}{section}

\newenvironment{textequation}[1][0.8]
{\begin{equation}
\begin{aligned}
\begin{minipage}{#1\linewidth}}
{\end{minipage}
\end{aligned}
\end{equation}
\ignorespacesafterend}

\newcommand{\btext}{\begin{textequation}}
\newcommand{\etext}{\end{textequation}}

\def\hinweis{\@startsection{subsection}{2}%
 \z@{0.7\linespacing\@plus 0.5\linespacing}{0.7\linespacing}%
{\normalfont\itshape\indent}}

\newcounter{hours}\newcounter{minutes}
\newcommand{\printtime}{%
\setcounter{hours}{\time/60}%
\setcounter{minutes}{\time-\value{hours}*60}%
\ifthenelse{\value{minutes}<10}{\thehours :0\theminutes}{\thehours:\theminutes}}

\usepackage[german,english]{babel}
\usepackage{graphicx}
\RequirePackage{amsmath}
\RequirePackage{bm}
\RequirePackage{amssymb}
\RequirePackage{upref}
\RequirePackage{amsthm}
\RequirePackage{enumerate}
\RequirePackage{pb-diagram}
\RequirePackage{amsfonts}
\RequirePackage[mathscr]{eucal}
\RequirePackage{verbatim}
\RequirePackage{xr}
\RequirePackage{graphicx}
\usepackage{calc}
\usepackage{xspace}
\usepackage{dsfont}

\makeatletter
\RequirePackage{color}
\newcommand{\ann}[1]{\renewcommand{\@makefnmark}{\mbox{$^{\color{red}{\@thefnmark}}$}}%
\footnote {#1}}
\makeatother

\RequirePackage{upref}
\RequirePackage{amsthm}
\RequirePackage{enumerate}
\usepackage[mathscr]{eucal}

\usepackage{xr-hyper}

\usepackage{calc}

\newlength{\oddsidemarginlength}
\newlength{\topmarginlength}

\newcounter{numberoflines}
\newcounter{tempcc}
\oddsidemargin=\oddsidemarginlength
\evensidemargin=\oddsidemargin
\topmargin=\topmarginlength
\usepackage[colorlinks=true,linkcolor=blue,citecolor=blue,urlcolor=blue]{hyperref}  

\begin{document}

\flushbottom


\title[Quantum development of a  Cauchy hypersurface]{The quantum development of an asymptotically Euclidean Cauchy hypersurface}

\author{Claus Gerhardt}
\address{Ruprecht-Karls-Universit\"at, Institut f\"ur Angewandte Mathematik,
Im Neuenheimer Feld 205, 69120 Heidelberg, Germany}
\email{\href{mailto:gerhardt@math.uni-heidelberg.de}{gerhardt@math.uni-heidelberg.de}}
\urladdr{\href{http://www.math.uni-heidelberg.de/studinfo/gerhardt/}{http://www.math.uni-heidelberg.de/studinfo/gerhardt/}}

%
\subjclass[2000]{83,83C,83C45}
\keywords{quantization of gravity, quantum gravity,  gravitational wave, quantum development, Yang-Mills field, Gelfand triplet, eigendistributions}
\date{\today}
%


\begin{abstract} 
In our model of quantum gravity the quantum development of a Cauchy hypersurface is governed by a wave equation derived as the result of a canonical quantization process. To find physically interesting solutions of the wave equation we employ the separation of variables by considering  a temporal eigenvalue problem which has a complete countable set of eigenfunctions with positive eigenvalues and also a spatial eigenvalue problem which has a complete set of eigendistributions. Assuming that the Cauchy hypersurface is asymtotically Euclidean we prove that the temporal eigenvalues are also spatial eigenvalues and the product of corresponding eigenfunctions and eigendistributions, which will  be smooth functions with polynomial growth, are the physically interesting solutions of the wave equation. We consider these solutions to describe the quantum development of the Cauchy hypersurface.
\end{abstract}

\maketitle

\tableofcontents

\setcounter{section}{0}
\section{Introduction}
In general relativity the Cauchy development of a Cauchy hypersurface $\so$ is governed by the Einstein equations, where of course the second fundamental form of $\so$ has also to be specified.

 In the model of quantum gravity  we developed in a series of papers \cite{cg:qgravity,cg:uqtheory,cg:uqtheory2,cg:qgravity2, cg:uf2, cg:uf3} we pick a Cauchy hypersurface, which is then only considered to be a complete Riemannian manifold $(\so,g_{ij})$ of dimension $n\ge 3$, and define its quantum development to be described by special solutions of the wave equation
\begin{equation}\lae{1.1}
\begin{aligned}
&\frac1{32}\frac {n^2}{n-1}\Ddot u-(n-1)t^{2-\frac4n}\D u-\frac n2t^{2-\frac4n}Ru+\al_1\frac n8 t^{2-\frac4n}F_{ij}F^{ij}u\\
&+\al_2 \frac n4 t^{2-\frac4n}\ga_{ab}\s^{ij}\F^a_i\F^b_iu+\al_2\frac n2 m t^{2-\frac4n}V(\F)u+nt^2\Lam u=0,
\end{aligned}
\end{equation}
in a globally hyperbolic spacetime
\begin{equation}
Q=(0,\un)\times \so,
\end{equation}
\cf \cite{cg:uf2}. The preceding wave equation describes the interaction of a given complete Riemannian metric $g_{ij}$ in $\so$ with a  given Yang-Mills  and Higgs field; $R$ is the scalar curvature of $g_{ij}$, $V$ is the potential of the Higgs field, $\Lam$ a negative cosmological constant, $m$ a positive constant, $\al_1,\al_2$ are positive coupling constants and the other symbols should be self-evident. The existence of the time variable, and its range, is due to the quantization process.
\br
For the results and arguments in \cite{cg:uf2}  it was completely irrelevant that the values of the Higgs field $\F$ lie in a Lie algebra, i.e., $\F$ could also be just an arbitrary scalar field, or we could consider a Higgs field as well as an another arbitrary scalar field. Hence, let us stipulate that the Higgs field could also be just an arbitrary scalar field. 
\er
If $\so$ is compact we also proved a spectral resolution of equation \re{1.1} by first considering a stationary version of the hyperbolic equation, namely, the elliptic eigenvalue equation
\begin{equation}\lae{1.11}
\begin{aligned}
&-(n-1)\D v-\frac n2Rv+\al_1\frac n8 F_{ij}F^{ij}v\\
&+\al_2 \frac n4 \ga_{ab}\s^{ij}\F^a_i\F^b_iv+\al_2\frac n2 m V(\F)v=\mu v.
\end{aligned}
\end{equation}
It has countably many solutions $(v_i,\mu_i)$ such that
\begin{equation}
\mu_0<\mu_1\le \mu_2\le \cdots,
\end{equation}
\begin{equation}
\lim \mu_i=\un.
\end{equation}
Let $v$ be an eigenfunction with eigenvalue $\mu>0$, then we  looked at solutions of \re{1.1} of the form
\begin{equation}\lae{1.6}
u(x,t)=w(t) v(x).
\end{equation}
$u$ is then a solution of \re{1.1} provided $w$ satisfies the implicit eigenvalue equation
\begin{equation}\lae{1.15}
-\frac1{32}\frac{n^2}{n-1}\Ddot w-\mu t^{2-\frac4n}w-nt^2\Lam w=0,
\end{equation}
where $\Lam$ is the eigenvalue.

We proved in \cite{cg:uf2} that for any stationary eigenfunction  $v_j$ with positive eigenvalue $\mu_j$  there is  complete sequence of eigenfunctions $w_{ij}$ of the temporal  implicit eigenvalue problem such that the functions
\begin{equation}
u_{ij}(t,x)=w_{ij}(t)v_j(x)
\end{equation}
are solutions of the wave equation, \cf  also \cite[Section 6]{cg:qgravity2}. 

However, for non-compact Cauchy hypersurfaces one has to use a different approach in order to quantize the wave equation \re{1.1}. Let us first consider the temporal eigenvalue equation
\begin{equation}\lae{4.36.1}
-\frac1{32} \frac{n^2}{n-1}\Ddot w+n\abs\Lam t^2w=\lam t^{2-\frac4n}w
\end{equation}
in the Sobolev space
\begin{equation}
H^{1,2}_0(\R[*]_+).
\end{equation}
Here, 
\begin{equation}
\Lam<0
\end{equation}
is the cosmological constant.

The eigenvalue problem \re{4.36.1} can be solved by considering the generalized eigenvalue problem for the bilinear forms
\begin{equation}
B(w,\tilde w)=\int_{\R[*]_+}\{\frac 1{32}\frac{n^2}{n-1}\bar w'\tilde w'+n\abs\Lam t^2\bar w\tilde w\}
\end{equation}
and
\begin{equation}
K(w,\tilde w)=\int_{\R[*]_+}t^{2-\frac4n}\bar w\tilde w
\end{equation}
in the Sobolev space $\mc H$ which is the completion of
\begin{equation}
C^\un_c(\R[*]_+,\Cc)
\end{equation}
in the norm defined by the first bilinear form.

We then look at the generalized eigenvalue problem
\begin{equation}\lae{4.43.1}
B(w,\f)=\lam K(w,\f)\q\A\,\f\in\mc H
\end{equation}
which is equivalent to \re{4.36.1}.
\bt\lat{4.1.1}
The eigenvalue problem \re{4.43.1} has countably many solutions $(w_i,\lam_i)$ such that
\begin{equation}\lae{4.44.1}
0<\lam_0<\lam_1<\lam_2<\cdots,
\end{equation}
\begin{equation}
\lim\lam_i=\un,
\end{equation}
and
\begin{equation}
K(w_i,w_j)=\de_{ij}.
\end{equation}
The $w_i$ are complete in $\mc H$ as well as in $L^2(\R[*]_+)$.
\et
Secondly, let $A$ be the elliptic operator on the left-hand side of \re{1.11}, assuming that its coefficients are smooth and bounded in any
\begin{equation}\lae{1.19}
C^m(\so),\qq m\in \N,
\end{equation}
then $A$ is self-adjoint in $L^2(\so,\Cc)$ and, if $\so$ is asymptotically Euclidean, i.e., if it satisfies the very mild conditions in \fras{3.1}, then the Schwartz space $\msc S$ of rapidly decreasing functions can also be defined in $\so$,
\begin{equation}
\msc S=\msc S (\so),
\end{equation}
such that
\begin{equation}
\msc S\su L^2(\so)\su\msc S'
\end{equation}
is a Gelfand triple and the eigenvalue problem in $\msc S'$
\begin{equation}
A f=\lam f
\end{equation}
has a solution for any $\lam\in \s(A)$, \cf \frt{2.4}. Let
\begin{equation}
(\msc E_\lam)_{\lam\in\s(A)}
\end{equation}
be the set of eigendistributions in $\msc S'$ satisfying
\begin{equation}
Af(\lam)=\lam f(\lam),\qq f(\lam)\in \msc E_\lam,
\end{equation}
then the $f(\lam)$ are actually smooth functions in $\so$ with polynomial growth, \cf \cite[Theorem 3]{cg:uf3}. Moreover, due to a result of Donnelly \cite{donnelly:spectrum2}, we know that
\begin{equation}
[0,\un)\su\s_{\tup{ess}}(A),
\end{equation}
hence, any temporal eigenvalue $\lam_i$ in \rt{4.1.1} is also a spatial eigenvalue of $A$ in $\msc S'$
\begin{equation}
Af(\lam_i)=\lam_if(\lam_i).
\end{equation}
Since the eigenspaces $\msc E_{\lam_i}$ are separable we deduce that  for each $i$ there is an at most countable basis of eigendistributions in $\msc E_{\lam_i}$
\begin{equation}
v_{ij}\equiv f_j(\lam_i),\qq 1\le j\le n(i)\le\un,
\end{equation}
satisfying
\begin{equation}
Av_{ij}=\lam_iv_{ij},
\end{equation}
\begin{equation}
v_{ij}\in C^\un(\so)\ii\msc S'(\so).
\end{equation}
The functions
\begin{equation}
u_{ij}=w_iv_{ij}
\end{equation}
are then smooth solutions of the wave equations. They are considered to describe the quantum development of the Cauchy hypersurface $\so$.

Let us summarize this result as a theorem: 
\bt
Let $A$ and $\so$ satisfy the conditions in \re{1.19} and  \ras{3.1}, and let $w_i$ \resp $v_{ij}$ be the countably many solutions of the temporal \resp spatial eigenvalue problems, then 
\begin{equation}
u_{ij}=w_iv_{ij}
\end{equation}
are smooth solutions  of the wave equation \re{1.1}. They describe the quantum development of the Cauchy hypersurface $\so$.
\et

\br
We used a similar approach  to describe the quantum development of the event horizon of an AdS blackhole, see \cite{cg:qbh}.
\er
\section{Existence of a complete set of eigendistributions} 

Let $H$ be a separable Hilbert space, $\mc S$ a complete nuclear space and
\begin{equation}\lae{2.1}
j:\mc S\hra H
\end{equation}
an embedding such that $j(\mc S)$ is dense in $H$. The triple
\begin{equation}
\mc  S\su H\su \mc S'
\end{equation}
is then called a Gelfand  triple and $H$ a \tit{rigged} Hilbert space. Moreover, we require that  the semi-norms $\norm\cdot_p$ defining the topology of $\mc S$ are a countable family.  In view of the assumption \re{2.1} at least of one the semi-norms is already a norm, since there exist a constant c and a semi-norm  $\norm\cdot_p$ such that
\begin{equation}
\norm{j(\f)}\le c \norm\f_p\qq\A\, \f\in\mc S,
\end{equation}
hence $\norm\cdot_p$ is a norm since $j$ is injective. But then there exists an equivalent sequence of norms generating the topology of $\mc S$. Since $\mc S$ is nuclear we may also assume that the norms are derived from a scalar product, \cf \cite[Theorem 2, p. 292]{ky}. 

Let $\mc S_p$ be the completion of $\mc S$ with respect to $\norm \cdot_p$, then 
\begin{equation}
\mc S= \iii_{p=1}^\un\mc S_p
\end{equation}
and
\begin{equation}\lae{2.5}
\mc S'=\uuu_{p=1}^\un\mc S_p'.
\end{equation}
A nuclear space $\mc S$ having these properties is  called a nuclear countably Hilbert space or a nuclear Fr\'echet Hilbert space.

Let $A$ be a self-adjoint operator in $H$ with spectrum
\begin{equation}
\Lam=\s(A).
\end{equation}
Identifying $\mc S$ with $j(\mc S)$ we assume
\begin{equation}\lae{2.7}
A(\mc S)\su \mc S
\end{equation}
and we want to prove that for any $\lam\in\Lam$ there exists
\begin{equation}
0\not=f(\lam)\in \mc S'
\end{equation}
satisfying
\begin{equation}
\spd{f(\lam)}{A\f}=\lam\spd{f(\lam)}\f\q\A\,\f\in\mc S.
\end{equation}
$f(\lam)$ is then called a generalized eigenvector, or an eigendistribution, if $\mc S'$ is a space of distributions. The crucial point is that we need to prove the existence of a generalized eigenvector for any $\lam\in\Lam$.

\bd
We define
\begin{equation}
\msc E_\lam=\set{f\in \msc S'}{Af=\lam f}
\end{equation}
to be the generalized eigenspace of $A$ with eigenvalue $\lam\in\Lam$ provided
\begin{equation}\lae{2.11}
\msc E_\lam\not=\{0\}.
\end{equation}
If \re{2.11} is valid for all $\lam\in\Lam$, then we call
\begin{equation}
(\msc E_\lam)_{\lam\in\Lam}
\end{equation}
a complete system of generalized eigenvectors of $A$ in $\mc S'$.
\ed

\bl\lal{2.2}
If $\mc S$ is separable, then each $\msc E_\lam\not= \{0\}$ is also separable in the inherited strong topology of $\mc S'$. 
\el
\bp
The  Hilbert spaces $\mc S_p$ are all separable by assumption, so are their duals $\mc S_p'$. Let $\mc B_p$ be a countable dense subset  of $\mc S_p'$ and set
\begin{equation}
\mc B=\uuu_{p=1}^\un \mc B_p,
\end{equation}
Then $\mc B$ is dense in $\mc S'$ in the strong topology. Indeed, consider $f\in\mc S'$ and a bounded subset $B\su \mc S$, then there exists $p$ such that $f\in \mc S'_p$, in view of \re{2.5}, and for any $g\in \mc B_p$ we obtain
\begin{equation}
\sup_{\f\in B}\abs{\spd {f-g}\f}\le\norm{f-g}_{-p}\sup_{\f\in B}\norm\f_p\le c_B\norm{f-g}_{-p}
\end{equation}
proving the claim. 
\ep
Let $E$ be the spectral measure of $A$ mapping Borel sets of $\Lam$ to projections in $H$, then we can find an at most countable family of mutually orthogonal unit vectors
\begin{equation}
v_i\in H,\qq1\le i\le m\le\un,
\end{equation}
and mutually orthogonal subspaces
\begin{equation}
H_i\in H
\end{equation}
which are generated by the vectors
\begin{equation}
E(\Om)v_i,\qq\Om\in \msc B(\Lam),
\end{equation}
where $\Om$ is an arbitrary Borel set in $\Lam$, such that
\begin{equation}
H=\bigoplus_{i=1}^mH_i.
\end{equation}
Each subspace $H_i$ is isomorphic to the function space
\begin{equation}
\hat H_i=L^2(\Lam,\Cc,\m_i)\equiv L^2(\Lam,\m_i),
\end{equation}
where $\mu_i$ is the positive  Borel measure
\begin{equation}
\mu_i=\spd{Ev_i}{v_i}.
\end{equation}
We have
\begin{equation}
\mu_i(\Lam)=1
\end{equation}
and there exists a unitary $U$ from $H_i$ onto $\hat H_i$ such that
\begin{equation}
\spd uv=\int_\Lam\bar{\hat u}(\lam)\hat v(\lam) d\mu_i \qq \A\, u,v\in H_i
\end{equation}
where we have set
\begin{equation}
\hat u= Uu\qq\A\, u\in H_i.
\end{equation}
Hence, there exists a unitary surjective  operator, also denoted by $U$,
\begin{equation}\lae{2.22}
U: H\ra \hat H=\bigoplus_{i=1}^m \hat H_i
\end{equation}
such that $u=(u^i)$ is mapped to
\begin{equation}
\hat u=Uu=(Uu^i)=(\hat u^i)
\end{equation}
and
\begin{equation}
\hat u^i=\hat u^i(\lam)\in L^2(\Lam, \mu_i).
\end{equation}
Moreover, if $u\in D(A)$, then
\begin{equation}\lae{2.25}
\wh{Au}=(\wh{Au}^i(\lam))=(\lam \hat u^i)=\lam \hat u.
\end{equation}
For a proof of these well-known results see e.g. \cite[Chap. I, Appendix, p. 127]{gelfand:book}.
\br
We define the positive measure
\begin{equation}
\mu=\sum_{i=1}^m2^{-i}\mu_i
\end{equation}
in $\Lam$, and we shall always have this measure in mind when referring to null sets in $\Lam$. Moreover, applying the Radon-Nikodym theorem, we conclude that there are non-negative Borel functions, which we express in the form $h_i^2, 0\le h_i,$ such that
\begin{equation}
h_i^2\in L^1(\Lam,\mu)
\end{equation}
and
\begin{equation}
d\mu_i=h_i^2d\mu.
\end{equation}
The map
\begin{equation}\lae{2.31}
v\in L^2(\Lam,\mu_i)\ra h_iv\in L^2(\Lam,\mu)
\end{equation}
is a unitary embedding.
\er 
\bl\lal{2.4}
The functions $h_i$ satisfy the following relations
\begin{equation}
\sum_{i=1}^m2^{-2i}h_i^2<\un \qq\mu \tup{ a.e.}
\end{equation}
and
\begin{equation}
\sum_{i=1}^m2^{-2i}h_i^2\not=0 \qq\mu \tup{ a.e.}
\end{equation}
Replacing the values of $h_i$ on the exceptional null sets by $2^{-i}$ the two previous relations are valid everywhere in $\Lam$.
\el
\bp
(i) We first prove that, for a fixed $i$,  $h_i$ cannot vanish on a Borel set $G$ with positive $\mu_i$ measure, $\mu_i(G)>0$. We argue by contradiction assuming that $h_i$ would vanish on  a Borel set $G$ with $\mu_i(G)>0$. Let $v\in H$ be arbitrary and let $v^i$ be the component belonging to $H_i$, then 
\begin{equation}
\begin{aligned}
\int_G\abs {\hat v^i}^2d\mu_i&=\int_\Lam\chi_G\abs{\hat v^i}^2 d\mu_i\\
&=\int_\Lam \chi_Gh_i^2\abs{\hat v^i}^2d\mu=0,
\end{aligned}
\end{equation}
and we deduce
\begin{equation}
\hat v^i=0\qq \mu_i\tup{ a.e. in }G\qq\A\, v\in H,
\end{equation}
a contradiction, since the $\hat v^i$ generate $L^2(\Lam, \mu_i)$.

\cvm
(ii) Now, let $G\su \Lam$ be an arbitrary Borel set satisfying $\mu(G)>0$ and define $\hat\psi=(\hat\psi^i)$ by setting
\begin{equation}
\hat\psi^i=\chi_G2^{-i}, 
\end{equation}
then we obtain
\begin{equation}
\begin{aligned}
\norm{\hat\psi}^2&=\sum_{i=1}^m\int_\Lam\chi_G2^{-2i}d\mu_i\\
&=\sum_{i=1}^m \int_\Lam \chi_G 2^{-2i}h_i^2d\mu<\un
\end{aligned}
\end{equation}
concluding
\begin{equation}
\sum_{i=1}^m2^{-2i}h_i^2<\un\qq\mu \tup{ a.e.}
\end{equation}
as well as
\begin{equation}
\sum_{i=1}^m2^{-2i}h_i^2\not=0\qq\mu \tup{ a.e.},
\end{equation}
where the last conclusion is due to the result proved in (i), since there must existst an $i$ such that $\mu_i(G)>0$.
\ep

Now we can prove:
\bt\lat{2.4}
Let $H$ be a separable rigged Hilbert space as above assuming that the nuclear space $\mc S$ is a Fr\'echet Hilbert space, and let $A$ be a self-adjoint operator in $H$ satisfying \re{2.7}. Then there exists a complete system of generalized eigenvectors $(\msc E_\lam)_{\lam\in\Lam}$. If $\mc S$ is separable, then each eigenspace $\msc E_\lam$ is separable.
\et
\bp
Since $\mc S$ is nuclear there exists a norm $\norm\cdot_p$ such that the embedding
\begin{equation}
j:\mc S_p\hra H
\end{equation}
is nuclear, i.e., we can write
\begin{equation}
j(\f)=\sum_{k=1}^\un\lam_k\spd{f_k}\f u_k\qq\A\, \f\in\mc S,
\end{equation}
where
\begin{equation}
0\le \lam_k\q\wed\q \sum_{k=1}^\un\lam_k<\un,
\end{equation}
\begin{equation}
f_k\in\mc S_p'\q\wed\q \norm{f_k}=1,
\end{equation}
and $u_k\in H$ is an orthonormal sequence. We may, and shall, also assume
\begin{equation}\lae{2.44}
u_k\in D(A),
\end{equation}
since $D(A)$ is dense in $H$: Let
\begin{equation}
v_k\in D(A)
\end{equation}
be a sequence of linearly independent vectors generating a dense subspace in $H$, then we can define an orthonormal basis $(\tilde v_k)$ in $H$ which spans the same subspace. Hence, there exists a unitary map $T$ such that
\begin{equation}
\tilde v_k=Tu_k\qq\A\, k\in\N.
\end{equation}
Instead of the embedding $j$ we can then consider the embedding
\begin{equation}
T\circ j                                       
\end{equation}
proving our claim. Thus, we shall assume \re{2.44} which is convenient but not necessary.

We immediately infer from the assumption that $j(\mc S)$ is dense in $H$ the following conclusions:
\btext
The $(u_k)$ are complete in H,
\etext
\begin{equation}
0<\lam_k\qq\A\,k,
\end{equation}
and
\btext\lae{2.33}
for all $k$ there exists $\f\in\mc S$ such that $\spd{f_k}\f\not=0$.
\etext

Let $U$ be the unitary operator in \re{2.22}, then we define
\begin{equation}
\hat\f=U\circ j(\f)=\sum_{k=1}^\un \lam_k\spd{f_k}\f\hat u_k
\end{equation}
such that
\begin{equation}
\hat u_k=(\hat u_k^i(\lam))_{1\le i\le m}
\end{equation}
\begin{equation}
\hat u_k^i\in L^2(\Lam,\mu_i).
\end{equation}
Applying the embedding in \re{2.31} we can also express $\hat u_k$ in the form
\begin{equation}
\hat u_k=(h_i\hat u^i_k(\lam))_{1\le i\le m}
\end{equation}
\begin{equation}
h_i\hat u^i_k\in L^2(\Lam,\mu).
\end{equation}

Similarly we have
\begin{equation}
\hat\f=(h_i\hat\f^i)
\end{equation}
and
\begin{equation}\lae{2.45}
\wh{A\f}=(\lam h_i\hat\f^i),
\end{equation}
in view of \re{2.25}. Here, we identify $\f$ and $j\f$, i.e.,
\begin{equation}
A\f\equiv A(j\f).
\end{equation}
We want to prove that
\begin{equation}\lae{2.59.1}
\begin{aligned}
\wh{A(j\f)}&=\sum_{k=1}^\un \lam_k\spd{f_k}\f\wh{A u_k}\\
&=\lam\sum_{k=1}^\un \lam_k\spd{f_k}\f\hat u_k.
\end{aligned}
\end{equation}
Indeed, for any bounded Borel set $\Om\su\Lam$
\begin{equation}
AE(\Om)
\end{equation}
is a self-adjoint bounded operator in $H$ such that
\begin{equation}
\norm{AE(\Om)}\le \sup_{\lam\in\Om}\abs\lam.
\end{equation}
Hence, we deduce
\begin{equation}
AE(\Om)(j\f)=\sum_{k=1}^\un \lam_k\spd{f_k}\f AE(\Om)u_k
\end{equation}
and
\begin{equation}
\wh{AE(\Om)u_k}=\lam \chi_\Om\hat u_k
\end{equation}
and we infer
\begin{equation}
\begin{aligned}
\chi_\Om\wh{A(j\f)}&=\chi_\Om\lam\sum_{k=1}^\un \lam_k\spd{f_k}\f\hat u_k\\
&=\chi_\Om\lam\hat\f.
\end{aligned}
\end{equation}
Since $\Om\su \Lam$ is an arbitrary bounded Borel set we conclude
\begin{equation}
\begin{aligned}
\wh{A(j\f)}&=\lam\sum_{k=1}^\un \lam_k\spd{f_k}\f\hat u_k\\
&=\lam\hat\f.
\end{aligned}
\end{equation}
The right-hand side of the second equation is square integrable and therefore the right-hand side of the first equation too.

Let us set
\begin{equation}
\hat\f(\lam)=(h_i\hat\f^i(\lam)).
\end{equation}
$h_i\hat\f^i$ is an equivalence class and to define $h_i\hat\f^i(\lam)$ as a complex number for a fixed $\lam\in\Lam$ requires to pick a representative of the equivalence class in order to define $h_i\hat\f^i(\lam)$. It is well-known that for a given representative $h_i\hat\f^i(\lam)$ is well defined for almost every $\lam\in\Lam$, i.e., apart from a null set. We shall show that $\hat\f(\lam)$ can be well defined for \tit{any} $\lam\in\Lam$ and any $\f\in\mc S$. The choices we shall have to make will be independent of $\f$.

Firstly, let us define the product
\begin{equation}
h_i\hat u_k^i
\end{equation}
unambiguously. In view of \rl{2.4} $h_i$ is everywhere finite, i.e., we only have to consider the case when $h_i=0$ and $\abs{\hat u_k^i}=\un$. In this case we stipulate that
\begin{equation}
h_i\hat u_k^i=0.
\end{equation}
This definition insures that the integrals, e.g.,
\begin{equation}
\int_\Lam \abs{h_i\hat u_k^i}^2d\mu
\end{equation}
will give the correct values, because of Lebesgue's monotone convergence theorem: approximate $\abs{\hat u_k^i}$ by
\begin{equation}
\min(\abs{\hat u_k^i},r),\qq r\in\N.
\end{equation}

Secondly, we observe that 
\begin{equation}
1=\norm{\hat u_k}^2=\sum_{i=1}^m\int_\Lam\abs{h_i\hat u_k^i(\lam)}^2,
\end{equation}
and hence
\begin{equation}\lae{2.40}
\sum_{i=1}^m\abs{h_i\hat u_i^k(\lam)}^2<\un\qq\tup{a.e. in }\Lam.
\end{equation}
Thirdly, we have
\begin{equation}\lae{2.41}
\sum_{k=1}^\un\sum_{i=1}^m\abs{h_i\hat u_k^i(\lam)}^2\not=0\qq\tup{a.e. in }\Lam.
\end{equation}
Indeed, suppose there were  a Borel set 
\begin{equation}
G\su \Lam
\end{equation}
such that 
\begin{equation}
0<\mu(G)=\sum_i2^{-i}\mu_i(G)
\end{equation}
and
\begin{equation}\lae{2.41.1}
\sum_{k=1}^\un\sum_{i=1}^m\abs{h_i\hat u_k^i(\lam)}^2=0\qq\tup{in }G,
\end{equation}
then there would exist $j$ such that
\begin{equation}
\mu_j(G)>0
\end{equation}
and we would deduce
\begin{equation}
0=\sum_{k=1}^\un \int_G\abs{h_j\hat u^j_k}^2d\mu=\sum_{k=1}^\un \int_G\abs{\hat u^j_k}^2d\mu_j,
\end{equation}
contradicting the fact that the $(\hat u^j_k)$ are a basis for $L^2(\Lam,\mu_j)$.

Fourthly, we have
\begin{equation}
\begin{aligned}
\sum_k\sum_i\int_\Lam\lam_k\abs{h_i\hat u_k^i(\lam)}^2d\mu=\sum_k\lam_k\norm{\hat u_k}^2=\sum_k\lam_k<\un,
\end{aligned}
\end{equation}
hence we deduce
\begin{equation}\lae{2.43}
\sum_k\sum_i\lam_k\abs{h_i\hat u_k^i(\lam)}^2<\un\qq \tup{a.e. in }\Lam.
\end{equation}
Now, for any $(i,k)$ we choose a particular representative of $h_i\hat u_k^i$ by first picking   the representative of $h_i$ we defined in \rl{2.4} and  a representative of $\hat u_k^i$ satisfying the relations in \re{2.40}, \re{2.41} and \re{2.43} and then defining the values of these particular representatives in the exceptional null sets occurring in the just mentioned relations by 
\begin{equation}
h_i=2^{-i}\q\wed\q \hat u_k^i=2^{-i}2^{-k}.
\end{equation}
Then $h_i\hat u_k^i(\lam)$ is well defined for any $\lam\in\Lam$ and the relations in \re{2.40}, \re{2.41} and \re{2.43} are valid for any $\lam\in\Lam$.

Moreover, the series
\begin{equation}
h_i\hat\f^i(\lam)=\sum_k\lam_k\spd{f_k}\f h_i\hat u_k^i(\lam)
\end{equation}
converges absolutely, since
\begin{equation}
\begin{aligned}
\sum_k\lam_k\abs{\spd{f_k}\f}\abs{h_i\hat u_k^i(\lam)}&\le \norm\f_p\sum_k\lam_k\abs{h_i\hat u_k^i(\lam)}\\
&\le \norm\f_p(\sum_k\lam_k)^\frac12(\sum_k\lam_k\abs{h_i\hat u_k^i(\lam)}^2)^\frac12<\un, 
\end{aligned}
\end{equation}
in view of \re{2.43}. 
\bd
Let us define the sequence space
\begin{equation}
l_2=\set{(a^i_k)}{\sum_k\sum_i\abs{a_k^i}^2<\un}
\end{equation}
with scalar product
\begin{equation}
\spd{(a^i_k)}{(b^i_k)}=\sum_k(\sum_i\bar a^i_kb^i_k).
\end{equation}
\ed
Thus, we have
\begin{equation}
(\lam_k\spd{f_k}\f h_i\hat u^i_k(\lam))\in l_2,
\end{equation}    
since
\begin{equation}
\lam_k^2<\lam_k 
\end{equation}     
for $k$ large. By a slight abuse of language we shall also call this sequence $\hat\f(\lam)$,
\begin{equation}
\hat\f(\lam)=(\lam_k\spd{f_k}\f h_i\hat u^i_k(\lam)).
\end{equation} 

We are now ready to complete the proof of the theorem. Let $\lam\in\Lam$ be arbitrary, then there exists a pair $(i_0,k_0)$ such that
\begin{equation}
h_{i_0}\hat u^{i_0}_{k_0}(\lam)\not=0,
\end{equation}   
in view of \re{2.41}, which is now valid for any $\lam\in\Lam$. Define
\begin{equation}
f(\lam)=(h_{i_0}\hat u^{i_0}_{k_0}(\lam))\in l_2
\end{equation}     
to be the sequence with just one non-trivial term. We may consider 
\begin{equation}
f(\lam)\in \mc S_p'\su \mc S'
\end{equation}          
by defining
\begin{equation}
\spd{f(\lam)}\f=\spd{f(\lam)}{\hat\f(\lam)}\qq\A\, \f\in \mc S,
\end{equation} 
where the right-hand side is the scalar product in $l_2$. Indeed, we obtain 
\begin{equation}
\begin{aligned}
\abs{\spd{f(\lam)}\f}&= \lam_{k_0}\abs{\spd{f_{k_0}}\f}\abs{h_{i_0}\hat u^{i_0}_{k_0}(\lam)}^2\\
&\le \lam_{k_0}\abs{h_{i_0}\hat u^{i_0}_{k_0}(\lam)}^2\norm\f_p^2\q\A\,\f\in\mc S
\end{aligned}
\end{equation}       
yielding
\begin{equation}
f(\lam)\in\mc S_p'.
\end{equation}    
Furthermore,
\begin{equation}
f(\lam)\not=0,
\end{equation}    
since there exists $\f\in\mc S$ such that 
\begin{equation}
\spd{f_{k_0}}\f\not=0,
\end{equation} 
in view of \re{2.33}.

$f(\lam)$ is also a generalized eigenvector of $A$ with eigenvalue $\lam$, since
\begin{equation}
\begin{aligned}
\spd{f(\lam)}{A\f}=\spd{f(\lam)}{\wh{A\f}(\lam)} =\spd{f(\lam)}{\lam\hat\f(\lam)}=\lam\spd{f(\lam)}\f
\end{aligned}
\end{equation}      
because of \re{2.45} and \re{2.59.1}.   The final conclusions are derived from \rl{2.2}.                     
\ep

\section{Properties of $\s(A)$ in the asymptotically Euclidean case}

Let $A$ be the elliptic operator
\begin{equation}\lae{3.1}
\begin{aligned}
&-(n-1)\D v-\frac n2Rv+\al_1\frac n8 F_{ij}F^{ij}v\\
&+\al_2 \frac n4 \ga_{ab}g^{ij}\F^a_i\F^b_iv+\al_2\frac n2 m V(\F)v.
\end{aligned}
\end{equation}
We want to prove that
\begin{equation}\lae{3.2}
[0,\un)\su \s(A),
\end{equation}
in order to be able to quantize the wave equation \fre{1.1}. Using the results in \cite{donnelly:spectrum2} we shall show that \re{3.2} or even the stronger result
\begin{equation}\lae{3.3} 
[0,\un)\su \s_{\tup{ess}}(A),
\end{equation}
where $\s_{\tup{ess}}(A)$ is the essential spectrum, is valid provided the following assumptions are satisfied:
\bas\laas{3.1}
We assume there exists a compact $K\su \so$ and a coordinate system $(x^i)$ covering $\so\sminus K$ such that $\so\sminus K$ is diffeomorphic with an exterior region
\begin{equation}
\Om\su \R[n]
\end{equation} 
and 
\begin{equation}
x=(x^i)\in \Om.
\end{equation}
The metric $(g_{ij})$ then has to satisfy
\begin{equation}\lae{3.6}
\lim_{\abs x\ra\un} g_{ij}(x)=\de_{ij},
\end{equation}
\begin{equation}\lae{3.7}
\lim_{\abs x\ra\un} g_{ij,k}(x)=0,
\end{equation}
where a comma indicates partial differentiation, and there is a constant $c$ such that
\begin{equation}\lae{3.8}
cr\le \abs x\le c^{-1} r\qq\A\,x\in\Om,
\end{equation}
where $r$ is the geometric distance to a base point $p\in K$.

Furthermore, we require that the lower order terms of $A$ vanish at infinity, i.e.,
\begin{equation}\lae{3.9}
\lim_{\abs x\ra\un}\{\abs R+\abs{F_{ij}F^{ij}}+\abs{\ga_{ab}g^{ij}\F^a_i\F^b_i}+\abs{V(\F)}\}=0.
\end{equation}
\eas
Let us refer the lower order terms with the symbol $V=V(x)$ such that
\begin{equation}\lae{3.10}
A=(n-1)\{-\D+V\},
\end{equation}
then we shall prove
\bt\lat{3.2}
The operator $A$ in \re{3.10} has the property
\begin{equation}
[0,\un)\su \s_{\tup{ess}}(A).
\end{equation}
\et
\bp
We first prove the result for the operator $(-\D+V)$. Let us define a positive  function
\begin{equation}
b\in C^\un(\so),
\end{equation}
such that
\begin{equation}
b(x)=\abs x\qq\A\, x\not\in B_R(p),
\end{equation}
where $B_R(p)$ is a large geodesic ball containing the compact set $K$. In view of the assumptions \re{3.6}, \re{3.7} and \re{3.8} $b$ satisfies the conditions $(i)$, $(ii)$ and $(iii)$ in \cite[Properties 2.1]{donnelly:spectrum2}. Moreover, the assumption \re{3.9}, which implies
\begin{equation}
\lim_{\abs x\ra\un}\abs V=0,
\end{equation}
insures that the condition $(iv)$ in \cite[Theorem 2.4]{donnelly:spectrum2} can be applied yielding
\begin{equation}
[0,\un)=\s_{\tup{ess}}(-\D+V).
\end{equation}
However, since only the inclusion
\begin{equation}\lae{3.16}
[0,\un)\su\s_{\tup{ess}}(-\D+V).
\end{equation}
is proved while the reverse inclusion is merely referred to, and we could not look at the given references, we shall only use \re{3.16}. This relation is proved by constructing, for each $\e>0$ and $\lam>0$, an infinite dimensional subspace $G_\e$ of $C^2_c(\so)$ such that
\begin{equation}
\int_M\abs{(-\D+V-\lam^2)v}^2\le \e^2 \int_M\abs v^2\qq\A\, v\in G_\e.
\end{equation}
Multiplying this inequality by $(n-1)^2$ we infer that \re{3.16} is also valid when the operator $(-\D+V)$ is replaced by
\begin{equation}
A=(n-1)(-\D+V)
\end{equation}
proving the theorem.
\ep
\section{The quantization of the wave equation}
The quantization of the hyperbolic equation \fre{1.1} will be achieved by splitting the equation into two equations: A temporal eigenvalue equation, an ODE,  and a spatial elliptic eigenvalue equation.

Let us first consider the temporal eigenvalue equation
\begin{equation}\lae{4.36}
-\frac1{32} \frac{n^2}{n-1}\Ddot w+n\abs\Lam t^2w=\lam t^{2-\frac4n}w
\end{equation}
in the Sobolev space
\begin{equation}
H^{1,2}_0(\R[*]_+).
\end{equation}
Here, 
\begin{equation}
\Lam<0
\end{equation}
is a cosmological constant.

The eigenvalue problem \re{4.36} can be solved by considering the generalized eigenvalue problem for the bilinear forms
\begin{equation}
B(w,\tilde w)=\int_{\R[*]_+}\{\frac 1{32}\frac{n^2}{n-1}\bar w'\tilde w'+n\abs\Lam t^2\bar w\tilde w\}
\end{equation}
and
\begin{equation}
K(w,\tilde w)=\int_{\R[*]_+}t^{2-\frac4n}\bar w\tilde w
\end{equation}
in the Sobolev space $\mc H$ which is the completion of
\begin{equation}
C^\un_c(\R[*]_+,\Cc)
\end{equation}
in the norm defined by the first bilinear form.

We then look at the generalized eigenvalue problem
\begin{equation}\lae{4.43}
B(w,\f)=\lam K(w,\f)\q\A\,\f\in\mc H
\end{equation}
which is equivalent to \re{4.36}.
\bt\lat{4.1}
The eigenvalue problem \re{4.43} has countably many solutions $(w_i,\lam_i)$ such that
\begin{equation}\lae{4.44}
0<\lam_0<\lam_1<\lam_2<\cdots,
\end{equation}
\begin{equation}
\lim\lam_i=\un,
\end{equation}
and
\begin{equation}
K(w_i,w_j)=\de_{ij}.
\end{equation}
The $w_i$ are complete in $\mc H$ as well as in $L^2(\R[*]_+)$.
\et
\bp
The quadratic form $K$ is compact with respect to the quadratic form $B$ as one can easily prove, \cf \cite[Lemma 6.8]{cg:qfriedman}, and hence a proof of the result, except for the strict inequalities in \re{4.44}, can be found in \cite[Theorem 1.6.3, p. 37]{cg:pdeII}. Each eigenvalue has multiplicity one since we have a linear ODE of order two and all solutions satisfy the boundary condition 
\begin{equation}\lae{6.45}
 w_i(0)=0.
\end{equation}
The kernel is two-dimensional and the condition \re{6.45} defines a one-dimen\-sional subspace. Note, that we considered only real valued solutions to apply this argument. 
\ep
The elliptic eigenvalue equation has the form
\begin{equation}\lae{4.12}
Av=\lam v,
\end{equation}
where $A$ is the elliptic operator in \fre{3.1} and $v\in C^\un(\so)$. $A$ is a self-adjoint operator in $L^2(\so,\Cc)$. Let 
\begin{equation}
\msc S=\msc S(\so)
\end{equation}
be the Schwartz space of rapidly decreasing smooth functions, then $\msc S$ is a separable nuclear Fr\'echet Hilbert space and
\begin{equation}
\msc S\su L^2(\so,\Cc)\su \msc S'
\end{equation}
a  Gelfand triple. Applying the results of \frt{2.4} we infer that there exists a complete system of eigendistributions
\begin{equation}
(\msc E_\lam)_{\lam\in\s(A))}
\end{equation}
in $\msc S'$, i.e.,
\begin{equation}
Af(\lam)=\lam f(\lam)\qq\A\, f(\lam)\in \msc E_\lam.
\end{equation}
These eigendistributions are actually smooth functions in $\so$ with polynomial growth as we proved in \cite[Theorem 3]{cg:uf3}. Assuming, furthermore, that the conditions in \fras{3.1} are satisfied we conclude that
\begin{equation}
[0,\un)\su\s_{\tup{ess}}(A),
\end{equation}
in view of \frt{3.2}, i.e., the equation \re{4.12} is valid for all $\lam\in \R[]_+$, and we  conclude further that each temporal eigenvalue $\lam_i$ of the equation \re{4.36} can also be looked at as a spatial eigenvalue of the equation \re{4.12}. Since the eigenspaces $\msc E_{\lam_i}$ are separable we deduce that for each $i$ there is an at most countable basis of eigendistributions in $\msc E_{\lam_i}$
\begin{equation}
v_{ij}\equiv f_j(\lam_i),\qq 1\le j\le n(i)\le\un,
\end{equation}
satisfying
\begin{equation}
Av_{ij}=\lam_iv_{ij},
\end{equation}
\begin{equation}
v_{ij}\in C^\un(\so)\ii\msc S'(\so).
\end{equation}
The functions
\begin{equation}
u_{ij}=w_iv_{ij}
\end{equation}
are then smooth solutions of the wave equations. They are considered to describe the quantum development of the Cauchy hypersurface $\so$.

Let us summarize this result as a theorem:
\bt
Let $\so$ satisfy the conditions in \ras{3.1} and let $w_i$ \resp $v_{ij}$ be the countably many solutions of the temporal \resp spatial eigenvalue problems, then 
\begin{equation}
u_{ij}=w_iv_{ij}
\end{equation}
are smooth solutions  of the wave equation. They describe the quantum development of the Cauchy hypersurface $\so$.
\et

\bibliographystyle{hamsplain}
\providecommand{\bysame}{\leavevmode\hbox to3em{\hrulefill}\thinspace}
\providecommand{\href}[2]{#2}



\end{document}